\documentclass{llncs}  
\usepackage{amsmath,epsfig,graphicx}
\usepackage{amsfonts}
\usepackage{amssymb}
\usepackage{latexsym}
\usepackage{graphicx}
\usepackage{comment}
\usepackage{meermacros}
\makeatletter

\newcommand{\eproof}{{\mbox{}\hfill\qed}\end{proof}}

\begin{document}

\title{Some structural complexity results for $\exists \R$}

\author{Klaus Meer
\and 
Adrian Wurm
\institute{Computer Science Institute, BTU Cottbus-Senftenberg\\ 
Platz der Deutschen Einheit 1\\
D-03046 Cottbus, Germany\\
\{meer,wurm\}@b-tu.de}}
\authorrunning{Klaus Meer and Adrian Wurm}

\maketitle

\begin{abstract}
The complexity class $\exr$, standing for the complexity of deciding the
existential first order theory of the reals as real closed field in the Turing model,
has raised considerable interest in recent years. It is well known that 
$\NP \subseteq \exr \subseteq {\rm PSPACE}.$ In their compendium \cite{Schaefer2024},
Schaefer, Cardinal, and Miltzow give a comprehensive presentation of results 
together with a rich collection of open problems.
Here, we answer some of them dealing with structural issues of $\exr$ as 
 a complexity class. We show analogues of the classical results of
Baker, Gill, and Solovay finding oracles which do and do not separate NP form $\exr$,
of Ladner's theorem showing the existence of problems in $\exr \setminus \NP$ not being complete for
$\exr$ (in case the two classes are different), as well as a characterization of $\exr$ by means 
of descriptive complexity.
\end{abstract}

\section{Introduction}
The existential theory of the reals collects all true sentences in existential first-order logic over $\R,$ i.e., sentences
of the form $\exists x_1 \ldots \exists x_n \Phi(x_1,\ldots,x_n),$ where $\Phi$ is a quan\-ti\-fier-free
formula expressing a Boolean combination of polynomial equalities and inequalities. The complexity
of deciding the truth of a given sentence plays a prominent role for  several computational models
and complexity classes. In the Blum-Shub-Smale model - henceforth BSS for short - 
over $\R$ basic arithmetic operations
can be executed with unit cost and the size of $\Phi$ is basically the algebraic size of a dense encoding
of all polynomials occuring in it. The above decision problem then turns out to be
$\NPR$-complete, where the latter class is the real number analogue of classical NP \cite{BSS}.
However, the decision problem also makes perfect sense in the Turing model
of computation if we restrict all coefficients for the polynomials in $\Phi$ to be rational. 
Then, its size is the usual bit-size and the problem of deciding truth is easily seen to be NP-hard.
In the Turing model,  the exact complexity of the existential theory for the reals is (wide) open.
Using deep results on quantifier elimination it has been shown that it can be decided within PSPACE,
see \cite{Canny} and also \cite{Renegar}. The open placement of the problem somewhere
between NP and PSPACE has led to the definition of an own complexity class denoted $\exr,$ see a 
precise definition in the next subsection.

Beside many other things, in \cite{Schaefer2024} the authors ask 
whether several  structural results  well known to hold for the 
classical class NP in the Turing model also
can be settled for $\exr.$ The latter include an analogue of the Baker-Gill-Solovay result \cite{BGS}
on existence of oracles separating $\exr$ from NP and the question
whether $\exr$ can be characterized by means
of descriptive complexity as done in \cite{Fagin} for NP. 
In this paper we give positive answers to these questions and also show that Ladner's result \cite{Ladner}
can be transformed showing that if we assume $\exr \neq \NP$ there exist problems in the set difference
not being $\exr$-complete.

A few words on the proofs might be appropriate. All above mentioned results for  NP turn
out to hold in the corresponding variants as well for $\exr$. In some cases, though not 
completely straightforward,
proofs follow from not too complicated modifications of known proofs of related results in the BSS model.
One basic ingredient for the proofs to work is the characterization of $\exr$ via
the BSS class BP$(\NPR^0)$ in Theorem \ref{satz:npr0=existsR} below together with the most important 
observation that the class of $\NPR^0$-machines in the BSS model is countable. 
Therefore, we see the main value of the present paper in working out the necessary proof details of
expected results in order to enlarge the body of structural results known to hold for $\exr.$

\subsection{Basic definitions: $\exr$ and BSS-machines without constants}

We now define precisely the class $\exr$ together with BSS computations using
only rational machine constants as well as related discrete complexity classes BP$(\pr^0)$
and BP$(\npr^0).$

The class $\exr$ was formally
introduced in \cite{Schaefer2009}, but somehow was examined already earlier, see  \cite{Schaefer2024} for a historic 
account on questions related to this class being studied already in the 1990s.
The class can most easily be defined starting from a complete problem and then building the downward 
closure under usual polynomial time many one reductions. In the meanwhile, several
problems have been shown to be complete. For our purposes, the existence of real solutions of a polynomial
system of degree-two equations is the most suitable one.

\begin{definition}\label{definition:qps-und-existr} 
a) \ The decision problem Quadratic Polynomial Systems $\QPSN$ is defined as follows:
Given $n,m \in \N$ and polynomials $f_i \in \Q[x_1,\ldots,x_m], 1 \leq i \leq n$
of degree at most $2$, is there an $x^* \in\R^n$ such that $f_i(x^*)=0$ for all $i?$
The problem is understood as a decision problem in the Turing model, i.e., instances
are coded as binary strings over $\{0,1\}^*$.\footnote{The superscript $^0$ in $\QPSN$ results from
the close relation of the approach to Blum-Shub-Smale machines that are only using
rational constants and is common in that context, see below. Here it indicates rational coefficients.}

b) \ The complexity class  $\exr$ consists of all decision problems $L \subseteq  \{0,1\}^*$
such that there exists a polynomial time many one reduction from $L$ to $\QPSN$.
\end{definition}

The completeness of $\QPSN$ for class $\exr$, i.e., for the existential theory of the reals easily follows 
from the proof in \cite{BSS} that the corresponding problem for quadratic polynomial systems with arbitrary
real coefficients is $\NPR$-complete in the BSS model over $\R$. In the corresponding 
proof, no real constants are introduced and starting from decision problems in $\{0,1\}^*$ the reduction 
runs in polynomial time also in the Turing model.

The class $\exr$ can easily be characterized via a restriction of the class $\NPR$ in the BSS model, 
see \cite{BSS,BCSS97}. The full real number BSS model allows
to compute with reals as entities, performing the basic arithmetic operations together with a test of the format
'is $x \geq 0$' at unit cost. Algorithms are allowed to use a fixed finite set of real numbers as 
so-called machine constants. The latter always include $0$ and $1.$ 
The (algebraic) size of an instance then is the number of reals necessary
to specify it (in a reasonable encoding), and the class $\PR$ of problems decidable in deterministic polynomial
time is defined literally the same as P, i.e., there exists a polynomial time algorithm (measured with respect
to the algebraic size measure for instances and the unit cost measure for
algorithms) deciding the problem correctly. Similarly, $\npr$ stands for problems verifiable in polynomial time 
when having access to a witness that is a real vector of polynomial (algebraic) size in the size of the given instance.
If we want to handle discrete problems in this model, i.e., problems $L \in \{0,1\}^*$, it is appropriate to
restrict inputs to binary strings and to disallow BSS algorithms to use non-rational machine constants. This leads to
complexity classes highly important in connection with $\exr$.

\begin{definition} The class {\rm BP}$(\PR^0)$ consists of all $L \subseteq  \{0,1\}^*$ such that
there exists a BSS algorithm running in (algebraic) polynomial time, using only rational machine constants
and deciding $L$ on inputs from  $\{0,1\}^*$. Similarly, the class  {\rm BP}$(\NPR^0)$
consists of all $L \subseteq  \{0,1\}^*$ such that
there exists a BSS algorithm using only rational machine constants
and verifying  $L$ on inputs from  $\{0,1\}^* \times \R^*$. Moreover, for inputs $(y,z)$ the 
(algebraic) running time is polynomially bounded in the size of $y.$  
\end{definition}

Above, BP stands for 'Boolean Part' and $\R^*$ denotes the real number analogue 
of $\{0,1\}^*$ , i.e., finite sequences
of real numbers. Note that for discrete inputs $y \in \{0,1\}^*$ the bit-size and the algebraic size
coincide.  However, the cost measure for algorithms working with discrete data is the 
algebraic one. It is also important to stress again  that for verification algorithms the certificate
$z \in \R^*$ is allowed to be a vector of reals. This is the reason for the tight relation of BP$(\NPR^0)$
to $\exr.$ It follows immediately from the original completeness proof in \cite{BSS}:

\begin{theorem}\label{satz:npr0=existsR}
{\rm BP}$(\NPR^0)\ = \  \exr.$ 
\end{theorem}

Below, a decisive aspect of this theorem is that $\exr$ can be characterized through
a complexity class defined by a machine model which has a countable number 
of machines only.

\section{Relativization of NP versus $\exr$}

The first 	question from the compendium \cite{Schaefer2024} we want to deal with is 
whether there exist oracles which do and do not
separate $\exr$ from NP and PSPACE, respectively.
Comparing P and NP in the Turing model, these questions were
answered affirmatively long ago \cite{BGS}, in the BSS model an analogue
result holds \cite{Emerson}, see also \cite{Gassner}. 
Usually, the construction of an oracle separating the representative classes is done
via a diagonalization argument. As consequence, the separating oracles in both settings
are not very natural (with an exception in \cite{Gassner}, where a Knapsack problem
is used to separate other classes). 
As we shall see, for NP and $\exr$ a quite natural oracle work. 
The definition of oracle classes like
P$^A$ and NP$^A$ naturally relies on a basic model for algorithms
defining a complexity class, like deterministic or non-deterministic polynomial time
Turing machines. Then, one equips them with the additional 
feature to use intermediate results as queries to an oracle. Following the same
nearby ongoing for $\exr$ we should work with the following definition.

\begin{definition}
Let $A \subseteq \R^*.$ A problem $L \subseteq \{0,1\}^*$ belongs to the oracle
class $\exr^A$, if there exists an $\NPR^0$-algorithm $M$ which in addition has an oracle state. 
If $M$ enters this state, it can ask the oracle whether an element $y \in \R^*$ 
previously computed belongs to $A$,  receives the correct answer in one step, and continues its
computation. Then, for $x \in L$ there
exists a computation of $M$ which accepts $x$ and for $x \in \{0,1\}^* \setminus L$ no
computation accepts.
\end{definition}

Note that though this is the straightforward definition of oracle classes $\exr^A$, some subtleties are hidden.
Since the basic machine is an $\NPR^0$-algorithm, it is allowed to produce real
number oracle questions, something a Turing machine cannot. Vice versa, since a Turing machine
only computes with discrete data, we know in advance that an NP-oracle machine will never
produce arbitrary non-rational queries, an information which is undecidable
for a BSS machine to know in general. Our results below establishing the integers
as separating oracle implicitly make use of such effects. We further comment on this at the
end of this section.

\begin{theorem}
Given the above definition of oracle classes based on $\exr$, the following hold:

a) \ There exists an oracle $A$ such that $\P^A = \NP^A = \exr^A = {\rm PSPACE}^A.$

b) \ The oracle $\Z$, seen as subset of $\R$, both satisfies  $\NP = \NP^{\Z} \subsetneq \exr^{\Z}$ and
$\exr^{\Z} \neq {\rm PSPACE}^{\Z}.$

c) \ In the full BSS model, $\PR^{\Z} \subsetneq \NPR^{\Z}.$  
\end{theorem}
\proof For a) we can choose $A$ as the PSPACE-complete problem Quantified Boolean Formulas
QBF. Since PSPACE = NPSPACE = co-NPSPACE it is well known that $\P^{\rm QBF} = 
\NP^{\rm QBF} = {\rm PSPACE}^{\rm QBF}= {\rm PSPACE}.$
Arguing about $\exr^{\rm QBF}$ needs some care since the underlying machine model in the definition 
changes.  Suppose $M$ to be a basic $\NPR^0$-oracle machine using
oracle QBF. Let $p(n)$ be the polynomial time running bound of $M$ for inputs in $\{0,1\}^n.$ W.l.o.g. suppose 
$M$ for such inputs asks $m \leq p(n)$ oracle queries, all being of polynomial size at most $p(n).$ 
A PSPACE-algorithm deciding $M^{\rm QBF} $ on input $x \in \{0,1\}^n$ works as follows: First, consider 
a computable enumeration of bit-vectors $b \in \{0,1\}^m$ and of $m$ strings $(q_1,\ldots,q_m)$, all $q_i$ strings 
over alphabet $ \{0,1,*\}$ and of size at most $p(n).$ Below, we interpret the $q_i$ as oracle queries $M$ poses during its computation 
and $b_i$ as the answers. A component $*$ in a query indicates that it does not belong to $\{0,1\}.$ 
In case that $M$ computes such a query, the answer from the QBF-oracle is $0$ since the query does not code a correct QBF-instance.  
Now, given a fixed tuple $(b_1,\ldots,b_m, q_1,\ldots,q_m)$ in the
enumeration, consider an existential first-order sentence over $\R$ without non-rational constants 
claiming that there exists a computation of $M$ on $x$ which produces the queries $q_i$ under the assumption that 
$b_1,\ldots, b_{i-1}$ are the correct answers to the previous queries, and then finally accepts. Since the $q_i$ are 
vectors over $\{0,1,*\},$
the requirement that a query computed by $M$ equals $q_i$ is easily expressible with such a first-order sentence. This holds for queries
being bit-vectors as well as for a query containing a component $*$, in which case the corresponding part of the first-order 
sentence has to express that the component does not belong to $\{0,1\}$. This is easily doable.
The resulting entire sentence is an instance of $\exr$ and thus can be decided in PSPACE by \cite{Canny}. It remains
to guarantee in addition that the $b_i$ are the correct answers to QBF-queries $q_i$. This of course can be done in PSPACE.
Note that if a $q_i$ contains a non-binary component, the corresponding $b_i$ must be $0$.
Now the algorithm does the above for all elements in the enumeration by re-using its space and accepts, if for at least one
tuple $(b_1,\ldots,b_m,q_1,\ldots,q_m)$ all conditions are satisfied.
It follows $\exr^{\rm QBF} = {\rm PSPACE}$ as well. 

For part b) note that both 
$\NP^{\Z}$ and ${\rm PSPACE}^{\Z}$ only contain  problems being decidable in the Turing model. Moreover,
if we consider a Turing machine over $\{0,1\}^*$  coding computations over $\Q$  in any usual way,
then it is easy to decide for an intermediate result whether it encodes an integer or not. Since for an NP-machine the size
of a query is polynomial in the input, we can also check in polynomial time whether a query is integral, thus
NP = NP$^{\Z}.$
But $\exr^{\Z}$ contains problems being undecidable in the Turing model such as Hilbert's 10th problem:
Given a polynomial $f \in \Z[x_1,\ldots,x_n]$ in $n$ variables with integer coefficients, an $\NPR^{0,\Z}$-algorithm
can guess in a non-deterministic computation $n$ real numbers $x_1^*,\ldots,x_n^*$ and then ask
the oracle whether all $x_i^*$ are integers. If the answer is positive the algorithm evaluates $f(x_1^*,\ldots,x_n^*)$
 in polynomial time in the algebraic model. It accepts iff the result is $0$. Thus, the problem to decide whether there is
 an integer zero of $f$ belongs to $\exr^{\Z},$ but is well known to be undecidable \cite{Matiyasevich}.

 For c) we can simplify Emerson's construction by using a result from \cite{Meer93}. Consider the
 set $A:=\{t \in [0,2\pi] | \exists k \in \N \ \mbox{s.t.} \  \frac{k\cdot t}{2 \pi} \in \N\}.$ It is easy to
 see that the decision problem: Given $x \in \R$, is $x \in A?$ belongs to $\NPR^{\Z}:$ For input
 $x$ guess a $ k \in \R$ and check whether $k\geq 0$ and $k\in \Z$ by asking the oracle. Then compute 
 $\frac{kx}{2\pi}$ and ask the oracle again whether it belongs to $\Z$. All this can be done in constantly
many algebraic steps. Note that in the BSS model $x\in \R$ has input size $1$. However, in order
to belong to $\PR^{\Z}$ the question should be decided in constant time using the oracle.  In \cite{Meer93}
it is shown by using a 'typical path argument' that this problem can not be decided 
by a polynomial time BSS machine which is allowed in addition to evaluate the sine-function in one step.
Since in this 'sine-model' one can decide in constant time whether a number is integral, it follows
can $\PR^{\Z}$ can be decided by a sine-machine, but $A$ can not. We conclude 
that $A \not\in \PR^{\Z}.$
\qed

\medskip
Note that part c) alternatively to the proof in \cite{Emerson} gives a more natural
problem  yielding the separation. As also for part b), this seems to be an effect 
observable at least for some
results in the BSS setting, compare a similar statement for a real number version 
of Post's problem \cite{MeerZiegler2005}.
A strange effect why the separation for NP and $\exr$ works is the fact that even 
though the former is a subset of the latter,
this does not any longer hold for arbitrary relativized versions. Clearly, the reason for
 this is the use of different
machine models to define the corresponding oracle classes. Such an effect concerning 
relativized classes is also known
from classical complexity theory for less prominent classes and studied in \cite{Vereshchagin}. 
It might be interesting to both find
other oracles yielding the separation and to study the above effect 
in more detail in our framework.


\section{Descriptive complexity for $\exr$}\label{section:dkt}

The next question from the compendium \cite{Schaefer2024} is whether $\exr$ can be
characterized by purely logical means in the sense of descriptive
complexity theory. Corresponding results were given in the early
years of finite model theory for classical NP \cite{Fagin} and later on for the
real number version $\NPR$ in the BSS model \cite{GraedelMeer}, based on 
the development of meta-finite model theory in \cite{GraedelGurevich}; see \cite{Graedeletal}
for a more intensive introduction into such results.
One special feature, given the above mentioned characterization of
$\exr$ as BP$(\NPR^0)$, is the mixture of discrete inputs coded as in classical
complexity theory over $\{0,1\}^*$, and an algebraic cost measure for the
arithmetic computations performed by the underlying machine. It turns out that 
this split can be modelled by a restriction of so-called $\R$-structures 
used in \cite{GraedelMeer} to characterize (full) $\npr$ and respective logics
on them. Thus, the focus below will be on defining the corresponding restriction
which we call \emph{discrete $\R$-structures}, and the suitable logics
on them used to capture both $\exr$ and BP$(\PR^0)$. Corresponding proofs
then are quite similar to those given in \cite{GraedelMeer} and are only sketched in the Appendix.
Though below we work with restricted $\R$-structures, we have to recall their
definition in full generality, except for disallowing arbitrary real constants in the 
so-called secondary part due to the fact that the BSS algorithms studied here do not use
such constants.

\begin{definition}\label{R-structure}
Let $L_s,L_f$ be finite vocabularies, where $L_s$ may
contain relation and function symbols,
and $L_f$ contains function symbols only.
An {\em $\R$-structure of signature}\/ $\sigma=(L_s,L_f)$ is a triple 
$\DD=(\cA,{\cal{R}}, \cal F)$ consisting of
\begin{description}
\item[(i)] 
a finite structure $\cA$ of vocabulary $L_s$, 
called the {\em primary part} or \emph{skeleton}\/ of $\DD$; its universe $A$   
is also said to be the {\em universe}\/ of $\DD$;
\item[ii)] the infinite structure ${\cal{R}} =(\R,0,1,+,-, \cdot, /, \sg,<)$ called the 
\emph{secondary part}. Here, $\sg :\R \mapsto \{0,1\}$ denotes the sign-function being $0$
for negative values and $1$ otherwise;
\item[(iii)]
a finite set $\cal F $ of functions $X: A^k\to\R$ interpreting the function
symbols in $L_f$, where $k$ depends on the respective symbol only.
\end{description}
We denote the set of all $\R$-structures of signature $\sigma$ 
by $\struc(\sigma)$.

For  $\DD \in \struc(\sigma)$ the {\em size} of $\DD$ is $|\DD| := |A|.$ 
\end{definition}

We now restrict $\R$-structures to so-called \emph{discrete $\R$-structures} suitable
for modelling computations in
$\pr^0$ and $\NPR^0$ on Boolean languages. In order to model bit-strings as structures
the universe always will have the form $A=\{0,1,\ldots,n-1\}$  and $L_s$ contains
a unary relation $X\subseteq A$ interpreted as a bit-vector in $\{0,1\}^n$ via
$X(i) = 1 \leftrightarrow i \in X,
X(i) = 0 \leftrightarrow i \not\in X.$ In addition, for well known
reasons we include a linear order $< \in L_s$ in order to capture polynomial time
computations below. Finally, as the only element in $L_f$ we need 
a function $real : A \mapsto \R$ which is able to change the type of an element
in $A$ to become a real (like the cost operator in a programming
language like $C_{++}):$  We define $ real(i) := 
\left\{ \begin{array}{ll} 1 \in \R & \mbox{for} \ i \neq 0, i \in A \\
0 \in \R & \mbox{for} \ i = 0 \in A \end{array} \right.$

\begin{definition} \label{definition:discrete-R-structure}
A \emph{discrete $\R$-structure} $\DD$ is an $\R$-structure over vocabulary
$\sigma = (L_s,L_f),$ where $A = \{0,1,\ldots,n-1\}$ for some $n \in \N, L_s$ contains
at least a unary relation $X$   and a binary relation $<$ interpreted as bit-string and linear order, 
respectively, and $L_f = \{real\},$ where $real : A \mapsto \R$ is interpreted as above.

We denote by $\strucn(\sigma)$ all discrete $\R$-structures with vocabulary $\sigma.$
\end{definition}
  
  Note that below when we use existential second-order (so) logic,  both
  $<$ and $real$   can be avoided by claiming their existence as functions from
  $A$ to $\R$ using an existential quantifier together with a fo-formula.

Discrete $\R$-structures and logics on them provide the right framework to model
BSS-computations on Boolean languages that do not use constants 
other than rational ones.
Inputs being discrete and measured by their usual bit size, the used logics will
transfer those inputs into the real number part of a discrete $\R$-structure
so that computations are modelled therein, including the use of the
algebraic unit cost measure.
Let $V= \{v_0,v_1,\ldots \}$ denote a countable set of variables. The $v_i$
are supposed to be fo-variables ranging over the discrete universe of a discrete
$\R$-structure.

\begin{definition}
Fix ${\cal{R}} = (\R,0,1,+,-,\cdot,/,\sg, <).$ 
The language $\FO^0$ contains, for each signature $\sigma=(L_s,L_f),$
a set of formulas and terms. Each term $t$ takes, when 
interpreted in some discrete $\R$-structure, values
in either the skeleton, in which case we call it an {\em index term},
or in $\R$, in which case we call it a {\em number term}.
Terms are defined inductively 
as follows
\begin{description} 
\item[(i)] 
The set of index terms is the closure of the set $V$ of variables
under applications of function symbols of $L_s$. 
\item[(ii)] Any rational number is a number term. 
\item[(iii)] 
If $h_1,\ldots,h_k$ are index terms and $X$ is a $k$-ary function
symbol of $L_f,$ then $X(h_1,\ldots,h_k)$ is a number term.
In particular, for $h$ an index term, $real(h)$ is a number term with value
in $\{0,1\} \subset \R.$ 

\item[(iv)] 
If $t,t'$ are number terms, then so are $t+t'$, $t-t'$, $t\cdot t'$, 
and $\sg(t)$. Here, the sign function is defined as 
$\sg(t) := \left\{ \begin{array}{ll} 1 \ & \ \mbox{if} \ t \geq 0 \\ 0 \ & \ \mbox{if} \ t < 0 \end{array} \ . \right.$
\end{description}

Atomic formulas are equalities $h_1=h_2$ of index terms, equalities
$t_1=t_2$ and inequalities $t_1<t_2, t_1 \leq t_2$ 
of number terms, and expressions $P(h_1,\ldots, h_k),$ where $P$ is
a $k$-ary predicate symbol in $L_s$ and $h_1,\ldots,h_k$ are index
terms.

The set of formulas of $\FO^0$ is the smallest set containing all
atomic formulas and which is closed under Boolean connectives and
quantification $(\exists v)\psi$ and $(\forall v)\psi$.  
\end{definition}

\begin{remark}
Having a linear order available on $A$ it is folklore in finite
model theory to define the first and the last element (denoted by $0$ and 
$n-1$, respectively)
of $A$ with respect to the order, as well as a linear order on every $A^k, k \in \N$
by means of fo-formulas. This is tacitly used below.
\end{remark}

In order to capture complexity class $\exr = {\rm BP}(\npr^0)$ as well as
BP$(\pr^0)$ we have to extend fo-logic to existential second-order logic and
first order  logic, respectively. This is done similarly
as in \cite{GraedelMeer} and the proofs showing that the resulting logics capture the intended
classes is a nearby variation of the corresponding ones in the full BSS model.
The only technical aspect to respect is that for discrete $\R$-structures
and the logics used we cannot introduce arbitrary real constants. This is rather 
a restriction for fixed point logic than for existential so-logic since for the
latter we can existentially quantify such real objects and then express 
their desired properties via a fo-formula. 
Since fixed point logic also needs to work with so-objects, we first define existential
so logic.

\begin{definition}
Second-order logic $\SO^0$ on discrete $\R$-structures  is obtained starting from $\FO^0$ logic by adding the possibility
to quantify over function symbols. More precisely, given a vocabulary $\sigma = (L_s,L_f),$ 
where $L_f$ contains a function
symbol $Y$ interpreted as a function from some $A^k \to \R$, together with a first-order formula 
$\phi$ over $\sigma$, both $\exists Y \phi(Y)$ and $\forall Y \phi(Y)$ are  second-order formulas.
If all quantified function symbols are existentially quantified we get {\em existential second-order logic} $\exists\SO^0.$
\end{definition}

\begin{example}
We express the core problem of $\exr$, namely real solvability of an instance of
QPS$^0$ (Definition \ref{definition:qps-und-existr}),
as  an existential so-property on suitable discrete $\R$-structures. Some (easy) technical
aspects are only sketched. Consider as input instance a system of $m \in \N$ polynomials
$p_1,\ldots,p_m \in \Z[x_1,\ldots,x_n]$ in some $n \in \N$ real variables having integer
coefficients. All $p_i$ are supposed to have degree at most $2$ and the question is to decide
whether there exists a common real zero. Recall that this is an $\exr$-complete problem.
Rational coefficients for sake of easiness in the further description can be
removed by multiplying all $p_i$ by the least common multiple of the denominators
of their respective coefficients.
We represent the system as follows as a discrete $\R$-structure $\DD=(\cA,{\cal{R}}, \{real\}),$
where $\cA$ has  the discrete set $A =\{1,\ldots,n\} \times \{1,\ldots,m\} \times \{0,\ldots,L\}$
as universe. Here, $L \in \N$ is an upper bound for the bit-size of the coefficients
of all $p_I.$ The vocabulary $L_s$ contains  a linear ordering
and a coefficient function $C:A^4 \mapsto \{0,1\} \subset A$, where for $i,j \in \{1,\ldots,n\},
k \in \{1,\ldots,m\}$ and $r \in \{0,\ldots,L\}$ we interpret $C(i,j,k,0)$ as sign of the coefficient 
of monomial $x_ix_j$ of $p_k;$ here $C(i,j,k,0) =0 \in A$ codes sign $-1$ and
$C(i,j,k,0) = 1 \in A$ codes sign $1.$
Furthermore, $C(i,j,k,r)$ is the bit (as element in $A$) of
$2^{r-1} $ in the binary representation of the coefficient of $x_ix_j$ in $p_k$.
Some technical comments are necessary. Formally, the above discrete part looks 
different from the definition of a discrete $\R$-structure.
However, it is an easy task to change the above structure to one with
 a universe $\tilde{A} = \{0,1,\ldots, K-1\}$ for some $K \in \N$ and
an $\tilde{X} \subset \tilde{A}$ interpreted as bit-string  coding the instance.
Towards this aim, one has to use an additional relation coding the values $n,m$ and $L$
in $\tilde{A}.$ Using the linear order on $\tilde{A}$ it is easy to express via fo-logic components
 representing  variables, polynomials, and coefficients, respectively.
Similarly, for a function $\tilde{C} : \tilde{A}^4 \mapsto
\{0,1\} \subset \tilde{A}$ it is easy to describe the correct arguments and their meaning
using fo-logic to define $\tilde{C}$ arbitrarily on arguments having no 
interpretation. The interested reader might try to elaborate this
coding; for sake of readability we prefer to use the above coding instead.

In the existential second order part we now ask for the existence of three functions
mapping some $A^k$ to $\R.$ First, $C_{\R} : A^4 \mapsto \R$ should represent
the coefficients as real numbers in order to use them in number terms
for expressing  the evaluation of the $p_k$ in real arguments. This can be 
done using the following fo-formulas. Note that properties like
$i,j \in \{1,\ldots,n\}, k \in \{1,\ldots,m\},
r \in \{0,\ldots, L\}$ can be expressed using the linear ordering; similarly,
expressions like $r+1$ can be defined to have the natural meaning.

First, the correct sign of a coefficient is expressed as
\[ \begin{array}{rcl}
\forall i,j \in \{1,\ldots,n\}, k \in \{1,\ldots,m\} \ C(i,j,k,0) =0 & \Leftrightarrow & 
C_{\R}(i,j,k,0) = - 1 \in \R \\
& & \hfill{(\mbox{as real number term})} \\
\wedge \ C(i,j,k,0) = 1 & \Leftrightarrow & C_{\R}(i,j,k,0) = 1 \in \R
\end{array}
\]
Next, $\forall i,j \in \{1,\ldots,n\}, k \in \{1,\ldots,m\} , r \in \{1,\ldots,L-1\}$ fixing the highest 
bit as real and describing the real represented by bit vector $C(i,j,k,*)$  is done via
\[ \begin{array}{l} C_{\R}(i,j,k,L) = real(C(i,j,k,L)) \ \mbox{and} \\ 
C_{\R}(i,j,k,r) =  C_{\R}(i,j,k,r+1) \cdot 2 + real(C(i,j,k,r)),
\end{array}
\]
where of course $2$ is a real number term for $ 1+1$. This way $C_{\R}(i,j,k,1)$ represents (as real)
the absolute value of the coefficient of monomial $x_i \cdot x_j$ in $p_k$.
Finally, we require for all $i,j,k : C_{\R}(i,j,k,L+1) = C_{\R}(i,j,k,0) \cdot C_{\R}(i,j,k,1)$ 
to  get the correct sign.

Next, use $\exists Y : A \mapsto \R$ in order to existentially
quantify a potential common real zero $(Y(1),\ldots, Y(n))$ of all $p_k.$ Finally,
we have to express the evaluation of all $p_k$ in $Y$.
However, this is easy and can be done similarly to the foregoing
construction. Existentially quantify a function $Z: A^3 \mapsto \R$
which represents the partial sums when evaluating all polynomials in $Y$. One component
in the arguments of $Z$ again addresses the respective polynomial to evaluate, the two
others code $x_i \cdot x_j.$ Since all $p_k$ have degree at most $2$, we need for
every $k\in \{1,\ldots,m\}$ to cycle through $\{1,\ldots,n\}^2$, and this can be expressed
similarly to the computation of a coefficient from a binary representation above.
The proceeding is the same as that in \cite{GraedelMeer} for evaluating a single degree-4 polynomial.
\end{example}

A logical characterization of $\exr$ now is possible in almost the same way as for $\npr$ in \cite{GraedelMeer}.
\begin{theorem}\label{satz:dkt-fuer-exr}
Let $(F,F^+)$ be a decision problem of discrete $\R$-structures, i.e., $F$ is a set of such structures, where
instance structures are coded as strings in $\{0,1\}^*$ as explained above, and $F^+ \subset F.$
Then $(F,F^+) \in \exr$ if and only if there is
an existential second-order sentence $\psi$ such that $F^+ = \{\DD \in F | \DD \models \psi\}.$
\end{theorem}

Dealing with a logical characterization of BP$(\PR^0)$ this is not possible, the
reason why at some places a bit more care for technical details is necessary. Nevertheless,
this is at most tedious.
The extension of FO$_{\R}^0$ to fixed point logic necessary in order
to capture polynomial time in our setting once again was basically introduced 
already in \cite{GraedelMeer}, here we outline the technical differences due to the lack
of arbitrary real constants.
The only new technical demand in order to describe $\P_{\R}^0$-computations using a logic
on discrete $\R$-structures 
is to change at certain places discrete data (i.e., objects formalized in the 
discrete part of an $\R$-structure) to real numbers. 
Suppose, for example, we use the given order on some $A^k$ and 
express by FO$_{\R}$ logic that $ \underline{t} \in A^k$  is the successor of $\underline{s}$ 
in this order, i.e., we have a formula saying $ \underline{t} =  \underline{s} +1. $ We then often want
to use the characteristic value $\chi [ \underline{t}  = \underline{s} +1]$ as real number. Formally, 
this can be expressed in our logics
as follows: $\underline{t}  = \underline{s} +1$ is an abbreviation of a formula
on the discrete part of an $\R$-structure. We can introduce a new discrete variable
$\sigma$ together with the fo-formula $\sigma  \Leftrightarrow \underline{t}  = \underline{s} +1.$
Now, $real(\sigma)$ is a number term expressing the truth value of 
$\underline{t}  = \underline{s} +1$ as real (!) number $0$ or $1$. We
express this by writing $real([\chi [ \underline{t}  = \underline{s} +1]).$
Similar constructions are used below and should be clear from the context.

We need the following definitions, compare \cite{GraedelMeer}.
In order to deal with partially defined functions from some $A^k \to \R$, enlarge
$\R$ by an element $undef$ and extend the arithmetic operations via
$r + undef := r - undef := undef , \\
r \cdot undef := r / undef := 
\left\{ \begin{array}{cc}
0 & r = 0\\
undef & r \neq 0
\end{array} \right.
$ and $ sgn(undef) := undef.$
For a number term $F( \underline{s}, \underline{t})$ with free first order variables 
$ \underline{s}, \underline{t}$, the $\max$ operator $\max\limits_{\underline{s}} F( \underline{s}, \underline{t})$
 is a number term with free variables  $\underline{t}$ and obvious semantics. Below, we
 use $\max$ in order to cycle through some $A^k.$
 Typically, we want $ \underline{s} $ to be the direct predecessor of $ \underline{t},$ then express
 the characteristic value $\chi[\underline{t} = \underline{s} +1] $ via FO$^0_{\R}$-logic,
 take the corresponding number term $ real(\chi[\underline{t} = \underline{s} +1] )$ as explained
 above, and then identify $ \underline{s}$ via $\max\limits_{\underline{s}} real(\chi[\underline{t} = \underline{s} +1] ).$
  That way, when a computation is in step  $ \underline{s}$ we can formalize the next step  $ \underline{t}.$
   In order to define fixed point logic on discrete $\R$-structures $\DD =(L_s, \cR, \{real\})$ let $Z$
  be a function symbol of arity $r$ interpreted on $\DD$ as function from $A^r \to \R$ and let
  $F(Z, \underline{x})$ be a number term in FO$^0_{\R} + \max$ logic
  on signature $(L_s, \{real, Z\})$ (i.e., the closure of FO$^0_{\R}$ logic under use of the $\max$ operator) 
  with free first order variables  $ \underline{x} =(x_1,\ldots,x_r) \in V^r.$
  
  \begin{definition}
 The fixed point $Z^{\infty}$ with respect to $F(Z,\underline{x})$ is a function
 symbol of arity $r.$ Its interpretation on a discrete $\R$-structure $\DD$
is defined iteratively as follows:
 Set $Z^{(0)}(\underline{x}) := undef \ \forall \underline{x} \in A^r$ and
 for $i \in \N : \\ Z^{(i)}(\underline{x}) := 
 \left\{ 
 \begin{array}{rl} 
  Z^{(i-1)}(\underline{x}) & \mbox{if it is defined already} \\
  F(Z^{(i-1)}, \underline{x}) & \mbox{if} \ Z^{(i-1)}(\underline{x}) = undef
 \end{array}.
 \right. $

 After at most $j \leq |A|^r$ iterations this process becomes
 saturated, i.e., $Z^{(j)} = Z^{(j+1)}.$ This fixed point
 then is $Z^{\infty}.$
  \end{definition}

\begin{definition}
Functional fixed-point logic $\FFP^0$ is obtained as closure of the set of first-order number 
terms under the maximization rule and the fixed-point rule. 
$\FP^0$ denotes the class of 
characteristic functions definable in functional fixed-point logic.
\end{definition}

\begin{theorem}\label{satz:dkt-fuer-prn}
On ranked discrete $\R$-structures the functions in $\FFP^0$ are exactly those computable 
in polynomial time in the constant-free BSS model. In particular, the characteristic 
functions in $\FFP^0$ are those of polynomial time solvable 
decision problems in $\{0,1\}^*,$ i.e. $\FP^0 ={\rm BP}(\PR^0).$ 
\end{theorem}

\smallskip
We finally note that another approach to characterize a certain fragment of $\exr$ by means
of a so-called probabilistic independence logic was given in \cite{Hannula}.

\section{A Ladner like theorem for NP and $\exr$}

The next classical result which can be transformed relatively
straightforwardly to $\exr$ is Ladner's theorem \cite{Ladner} guaranteeing the existence
of intermediate problems between P and the NP-complete ones. The range of the classical 
proof technique has been analyzed  in \cite{Schoening}. In relation with the BSS model of computation over uncountable
structures the theorem has been shown in \cite{MalajovichMeer} to hold
as well over the complex numbers. Over $\R$ it is open, but \cite{BMM,ChapuisKoiran,Meer2012}
give partial results. Especially,  \cite{BMM} addresses the importance of quantifier elimination QE algorithms
when constructing intermediate problems. 
In the case of separating the class of $\exr$-complete problems from NP under the assumption
$\exr \neq \NP$ a combination of the arguments from \cite{BMM,MalajovichMeer} works
as well. 

\begin{theorem}\label{Satz:Ladner} Suppose $\exr \neq \NP.$ Then, there exists a problem in $\exr \setminus \NP$ which
is not $\exr$-complete.
\end{theorem}

For the rest of this section we outline the proof idea of how to construct such a problem.
Technical details then easily can be filled 
following the presentations  in \cite{BMM,MalajovichMeer}.
Our construction starts from the $\exr$-complete problem $\QPSN$ of Definition \ref{definition:qps-und-existr}.
 As mentioned before,  the $\NPR$-completeness proof of the corresponding problem
 for the class $\NPR$ in \cite{BSS} shows as well completeness of $\QPSN$ in $\exr$ since 
 given an  $\NPR^0$-machine the reduction to the full $\QPS$ problem, i.e., where polynomials
 are allowed to have real coefficients as well, does not introduce real constants and runs in
 polynomial time also in the Turing model; intermediate results can be subsumed 
 under real existential quantification.\footnote{Note that guessing a real solution might lead 
 to real intermediate results; they can be existentially quantified in  $\exr$ in the reduction proof.}

The main reasons why a proof of Theorem \ref{Satz:Ladner} can be done similarly to  
showing related results are the decidability of $\exr$ within a computable time bound in the Turing model due
to real quantifier elimination, and effective countability of NP-machines together with polynomial
time bounds.
Starting from $\QPSN$ we build certain restricted languages $L(a)$ depending 
on strictly increasing sequences $(a) := (a_i)_{i \in \N}$ of natural numbers. For input dimensions
$n \in \{a_{2i-1},\ldots,a_{2i}-1\}$ we let $L(a)$ be the $\QPSN$ problem, for the remaining input dimensions
$L(a)$ equals the empty set. The strategy now is to define two such sequences $(a), (b) \in \N^{\N}$
so that both have the desired properties. We need two sequences in order to guarantee non-completeness.

\begin{definition} [cf. \cite{MalajovichMeer}] \label{definition:L(a)}
Let  $(a), (b) \in \N^{\N}$ be two sequences of natural numbers.

a)\  The sequences have an \emph{exponential gap} iff both are strictly increasing and 
for all $i \in \N$ satisfy
$a_{2i+1} \geq  2^{b_{2i}}$ as well as $b_{2i+1} \geq 2^{a_{2i+2}},$i.e.,
$a_1 < a_2 < 2^{a_2} \leq b_1 < b_2 < 2^{b_2} \leq a_3 \ldots. $

b)\ The decision problem $L(a)$ (and $L(b)$ similarly) is defined dimension-wise for inputs of size $n$ as follows:
\[ L(a) \cap \{0,1\}^n := \left\{ 
\begin{array}{cl} \QPSN \cap \{0,1\}^n & \mbox{if} \ \exists i \in \N \ \mbox{s.t.} \ a_{2i-1} \leq n < a_{2i} \\
\emptyset & \mbox{otherwise}
\end{array},
\right.
\]
\end{definition}

Since $\exr$-completeness is defined downward from $\QPSN$ by usual polynomial
time reductions, it is easy to see that assuming $\exr \neq \NP$ neither $L(a)$ nor $L(b)$ can be $\exr$-complete
if $(a)$ and $(b)$ have an exponential gap.

\begin{lemma} \label{lemma:L(a)notcomplete}
Suppose $\exr \neq \NP$ and let sequences $(a), (b)$ have an exponential gap. Suppose furthermore that 
$L(a), L(b)$ belong to $\exr \setminus \NP.$ 
Then both are not $\exr$-complete.
\end{lemma}
\proof Suppose one problem, say $L(a),$ is complete. Then there exists a polynomial time 
reduction $f$ from $L(b)$ to $L(a).$ Given an input $x \in \{0,1\}^n$ for $L(b)$, there are two cases.
If $x$ is a positive $\QPSN$-instance, i.e., $n$ belongs to some $\{b_{2i-1},\ldots,b_{2i}-1\}$, then
$f(x)$ must be a positive instance for $L(a).$ This is only possible if its size
$|f(x)|$ is at most $a_{2i} \leq \log{n}.$ In this case we can solve the $\QPSN$-instance
$f(x)$ by an algorithm like \cite{Renegar,Canny}, which are PSPACE-algorithms and thus run in 
exponential time in $\log{n}$, i.e., in polynomial time in $n$. If the size of $f(x)$ is larger
than $\log{n}$, then $a_{2i} \leq n < b_{2i-1}$ and thus $x \not\in L(b).$
This gives a polynomial time decision algorithm for $L(b)$ and thus $L(b) \in \P \subseteq \NP$
contradicting our assumption.
\qed

\smallskip

The main task is to construct $(a)$ and $(b)$ so that $L(a)$ and
$L(b)$ belong
to $\exr \setminus \NP.$ Below, we define both sequences stepwise
in the order $a_1, a_2, b_1, b_2, a_3, a_4, b_3, \ldots$ The construction will
guarantee the following properties:

i)\ $b_{2i-1} \geq 2^{a_{2i}}, a_{2i+1} \geq 2^{b_{2i}} \ \forall i \in \N$;

ii)\ the following task is efficiently solvable: Given $n \in \N$ in unary notation,
compute the maximal $j,\ell$ such that $a_j \leq n$ and $b_{\ell} \leq n$;

iii)\ let $(M_i,p_i)_i$ be an effective enumeration of all $\NP$-machines running in polynomial time
($M_i$ is restricted to run in time $p_i$ for a polynomial $p_i \ \Q[x]).$
Then for all $i \in \N$ there exists a $\QPSN$-instance $x \in \{0,1\}^n$ with
$n \in \{a_{2i+1}, \ldots, a_{2i+2}-1\}$ such that $M$ accepts $x$ iff $x$ is unsolvable, i.e., $x$ is an instance for which
$M_i$ does not work correctly in time $p_i(|x|).$  

\begin{lemma} \label{lemma:L(a)-in-exr}
If ii) above can be accomplished, then  $L(a)$ and $L(b)$
belong to $\exr.$
\end{lemma}
\proof A polynomial time reduction from $L(a)$ to $\QPSN$ works as follows: Given an $x \in \{0,1\}^n$,
compute the maximal $j$ and corresponding $a_j$ such that $a_j \leq n.$ If $ j = 2j'$ is even,
then $ n \in \{a_{2j'}, \ldots, a_{2j'+1}-1\}$ and thus $L(a) \cap \{0,1\}^n$ equals $\QPSN  \cap \{0,1\}^n.$
The reduction then outputs $x$. If  $ j = 2j'+1$ is odd,
then $L(a) \cap \{0,1\}^n = \emptyset$ and the reduction outputs a fixed 
unsolvable $\QPSN$ instance. Similarly for $L(b).$
\qed
 
 Now let $\{(M_i,p_i)| i \in \N \}$ be an effective enumeration of all NP-Turing machines with running time 
 bounded by a polynomial. W.l.o.g. we can assume for $M_i$ that it is a 
 non-deterministic Turing machine which is
 forced to stop its computation on $x$ after at most $p_i(|x|)$ steps and 
 rejects in case the computation has not been completed.
 We describe how to define sequences $(a)$ and $(b)$ such that  
i)-iii) above  are satisfied. The construction relies on known results that allow to describe
computations of both NP-algorithms and of $\NPR^0$-algorithms via existential 
first-order sentences over the real numbers.

\begin{proposition}
a) \ Let $\QPSN$ be as above and $n \in \N$. Then there exists an effectively computable existential 
fo-formula $\varphi^{(n)}$ in the theory of $\R$
as a real closed field with coefficients from $\Q$ only
such that $\forall x \in \{0,1\}^n : \ \varphi^{(n)}(x) \ \Leftrightarrow \ 
x$  is solvable as $\QPSN$-instance.

b) \ For every $\NP$-machine $M_i$ (in the Turing model) with
running time bounded by a polynomial $p_i$ and for every $n \in \N$
there exists an effectively computable fo-formula $\phi_i^{(n)}$ in the theory 
of $\R$ as a real closed field with coefficients from $\Q$ only
such that $\forall x \in \{0,1\}^n:  \ \phi_i^{(n)}(x) \ \Leftrightarrow \ $
there is an accepting computation of $M_i$ on
$x$ in  at most $p_i(|x|)$ steps.

In both statements, \emph{effectively computable} means computable in the Turing model
given the machines and $n$ using a time bound depending on the input size only.
\end{proposition}
\proof a) \ Let $M$ be the $\NPR^0$ BSS-machine for $\QPSN$ existing according to
Theorem \ref{satz:npr0=existsR}. Existence and  effective computability 
of $\varphi^{(n)}$ follows from the analogue statement
for general BSS-machines given in \cite{Michaux}, see also \cite{BMM}. Even the 
formula obtained for the latter result in the more general setting does not involve 
non-rational constants, so the same holds for $\varphi^{(n)}$ and its construction
can be performed by a Turing machine. Note that the variables
existentially quantified in $\varphi^{(n)}$ describe both the potential real solution of an instance and
the intermediate results of a computation of $M.$

b) \ Follows easily from a) by considering $M_i$ as an $\NPR^0$-algorithm.
A (discrete) guess $y$ of $M_i$ can  be expressed via existential quantification ranging over $\R$
by requiring the equation $y(y-1)=0$ to be fulfilled.
\qed

\smallskip

For given $i,n \in \N$  define another fo-formula $\Theta_i^{(n)}$ over $\R$ as
\[ \Theta_i^{(n)} \  \equiv \ \exists x \in \{0,1\}^n \ \neg (\varphi^{(n)}(x) \ \Leftrightarrow \ \phi_i^{(n)}(x)) \]
It states that there exists an instance $x \in \{0,1\}^n$ for which
NP-machine $M_i$ within $p_i(n)$ steps gives the false result w.r.t. $x$ as instance for
$\QPSN$, i.e., $(M_i,p_i)$ makes
an error concerning verifying $\QPSN$ on inputs of dimension $n$. Thus, if we define $(a)$ in such a way
that there is an $n = n(i)$ in $\{a_{2i-1}, \ldots a_{2i}-1\}$ such that $\Theta_i^{(n(i))}$ holds,
$L(a)$ will not be correctly decided by $M_i$ within
time bound $p_i$; and similarly for $L(b).$

We now describe an algorithm which at the same time defines and computes
$(a)$ and $(b)$, thereby proving condition ii) above to hold. Its structure
is such that given $n \in \N$ or later  an input $x \in \{0,1\}^n$, respectively, the 
computation of all $a_i, b_j < n$  is done anew.

\smallskip

{\bf Algorithm defining and computing $(a),(b)$}:
\begin{itemize}
\item[(1)] Define $a_1 := 0$.
\item[(2)] For given $ n > a_1$ do the following for every $k=a_1+1, \ldots, n:$
\item[(2.1)] Decide validity of $\Theta_k^{(1)}$ by using any of the quantifier elimination
algorithms mentioned above. If this algorithm needs more
than $n$ steps, then $a_2$ is not yet defined, but is guaranteed 
to satisfy $a_2 >n.$
\item[(2.2)] If a $k \leq n$ is found within at most $n$ steps such that $\Theta_k^{(1)}$ holds, then define $a_2 := n$ as well as
$b_1 := 2^{a_2}.$
\item[(3)] Compute similarly $b_2$ w.r.t. finding a $k$ with $b_1 \leq k < b_2$.
\end{itemize}

For a general $n$ suppose within at most $n$ steps numbers $a_1, \ldots, a_{2i+1},b_1,\ldots, b_{2i}$ 
have been computed. Then $a_{2i+2}$  is computed as follows for $n \geq a_{2i+1}:$

\begin{itemize}
\item[(4)]  Do the following for every $k=a_{2i+1}, \ldots, n:$
\item[(4.1)] Decide validity  of $\Theta_k^{(i)}$ by 
quantifier elimination.
If the execution of both the computation of $a_1,\ldots, b_{2i}$ and the
quantifier elimination exceeds $n$ steps, then $a_{2i+2}$ is not yet defined but guaranteed
to satisfy $n< a_{2i+2}.$
\item[(4.2)] If $k \leq n$ is found within $n$ steps such that $\Theta_k^{(i)}$ holds,
then define $a_{2i+2} := n$  as well as
$b_{2i+1} := 2^{a_{2i+2}}.$
\item[(5)] Compute $b_{2i+2}$ similarly.
\end{itemize}

\begin{theorem}
Assume $\exr \neq \NP$ and define sequences $(a) $ and $(b)$ according to the previous algorithm.
Then both $L(a)$ and $L(b)$ belong to $\exr \setminus \NP$, but none is $\exr$-complete.
\end{theorem}
\proof Given the way the algorithm defines $(a),(b)$, condition ii) above is satisfied, i.e., the assumption
of Lemma \ref{lemma:L(a)-in-exr} holds. It follows that both problems belong to $\exr.$
Furthermore, assuming $\exr \neq \NP$, the construction of both sequences implies
that no NP-machine $M$ with a polynomial time bound $p$ will verify $L(a)$ and
neither $L(b)$ correctly according to the requirements of being member in NP since
any such $(M,p)$ occurs in the effective enumeration as some $(M_i,p_i).$ 
By construction $(a) $ and $(b)$ have an exponential gap, so by Lemma \ref{lemma:L(a)notcomplete}
none of them is $\exr$-complete.
\hfill{ \ \ \qed}

\newpage
\section{Appendix}

In this Appendix we present the missing proof from Section \ref{section:dkt}

\proof (of Theorem \ref{satz:dkt-fuer-exr})
The main argument in order to end up in the waterway of proving
the analogue for $\npr$ is the earlier characterization of $\exr$ as BP$(\NPR^0).$
For the if-part, it is then straightforward to see that an $\npr^0$-computation
can guess the real valued function existentially quantified in $\psi$
and then evaluate in (algebraic) polynomial time the remaining
FO$_{\R}^0$-formula. Vice versa, let $L \subseteq \{0,1\}^*$ be a problem in 
$\exr = {\rm BP}(\npr^0)$ and let $M$ be a respective BSS-machine without non-rational constants
verifying $L$ in (algebraic) polynomial time. For an input $x \in \{0,1\}^n$ suppose
$M$ uses at most $n^m$ steps and registers for suitable $m$. If $x$ is represented
as discrete $\R$-structure $\DD = (\cA, \cR, \{real\})$ with universe
$A = \{0,\ldots, n-1\}, X   \subseteq A$ as described above, we can construct an FO$^0_{\R}$-formula
$\Phi$ on $(\DD,Y)$ for a $Y: A^{2m} \mapsto \R$ describing an accepting computation
of $M$ on $x$. Here, as usual in the classical proofs
for NP-completeness (Cook's theorem) and $\npr$-completeness (BSS theorem),
$Y$ encodes an entire computation table of $M$ of an accepting computation on $x$ such that
$x \in L \Leftrightarrow \exists Y : (\DD, Y) \models \Phi.$ The 
construction of $\Phi$ is done precisely as in \cite{GraedelMeer}, the only new
important aspect is to note that  $M$ is constant-free, so the resulting 
formula $\Phi$ belongs to $\exists SO_{\R}^0.$
\qed

\medskip

\proof (of Theorem \ref{satz:dkt-fuer-prn})
The proof is almost the same as the one in \cite{GraedelMeer} for characterizing
$\PR$ via fixed point logic on arbitrary $\R$-structures. For one direction, express the computation of 
a $\pr^0$-machine on an input from $\{0,1\}^*$ in time $\leq n^m$ via a
suitable function $Z : A^{2m} \to \R.$ The computation can easily be described 
by a number term 
$$ F(Z,\underline{j},\underline{t}) = \chi (\underline{t}=0) \cdot  F_{input}(\underline{j}) +
 \max\limits_{\underline{s}} \chi(\underline{t}=\underline{s}+1) \cdot 
 \sum\limits_{k=1}^N \chi(Z(0,\underline{s})=k) \cdot F_k(Z,\underline{j},\underline{s}),$$
where $\underline{j} \in A^m$ addresses a suitable register of the 
computation and $\underline{t} \in A^m$ a time step. This term stands in 
FO$_{\R}^0 + \max$ logic; moreover, for a given
discrete $\R$-structure $Z^{\infty}$ codes the entire computation and its 
value $Z^{\infty}(\underline{j}_{res}, |A|^m) \in \{0,1\}$ 
gives the correct decision for the question '$x \in L?$', where 
$\underline{j}_{res}$ is supposed to contain the final result of $M$. 
Vice versa, since the fixed point construction is guaranteed
to become stable after at most a polynomial number of steps -
only that many are necessary to cycle through some $A^k$ for fixed $k$ - every
formula in fixed point logic can be evaluated in polynomial
time in the constant-free BSS model for Boolean inputs. 
The details can be found in \cite{GraedelMeer} and can be
transferred almost literally to our setting. 
\qed

\end{document}